\documentclass{an}
\usepackage{times}
\usepackage{graphicx}
\begin{document}

\Yearpublication{2009}
\DOI{2009}
\title{An Instability in Triaxial Stellar Systems}
\author{Fabio Antonini\inst{1,2}
\and Roberto Capuzzo-Dolcetta\inst{1}
\and David Merritt\inst{2}
}

\institute{Department of Physics, La Sapienza, University of Rome
\and
Department of Physics and Center for Computational Relativity and Gravitation, 
Rochester Institute of Technology}

\abstract{The radial-orbit instability is a collective phenomenon that has 
heretofore only been observed in spherical systems. We find that this instability occurs also in triaxial
systems, as we checked by performing extensive $N$-body simulations whose initial conditions were
obtained by sampling a self-consistent triaxial model of a cuspy galaxy composed of luminous and dark
matter. $N$-body simulations show a time evolution of the galaxy that
is not due to the development of chaotic motions but, rather, to the collective instability induced by
an excess of box-like orbits. The instability quickly transforms such models into a more
prolate  configuration, with $0.64<b/a<0.77$ and $0.6<c/a<0.7$ for the dark halo and $0.64<b/a<0.77$  and $0.59<c/a<0.67$ for the
luminous matter. Stable triaxial, cuspy galaxies with dark matter halos are obtained when the contribution of radially-biased  orbits to the solution is reduced. These results constitute the first evidence of the
radial-orbit instability in triaxial galaxy models.}

\maketitle

\section{Introduction}
The aim of the current work is to test, by extensive $N$-body simulations, the stationarity of the triaxial galaxy
models realized by Capuzzo Dolcetta et al. (2007, hereafter CLMV07). 
These  models were constructed by  means of the standard orbital 
superposition method introduced by Schwarzschild (1979), and represent triaxial, cuspy elliptical galaxies embedded in triaxial dark halos 
in which the dark  and the luminous components have the same axial ratios 
($a=1, b=0.86, c=0.7$, i.e., maximal triaxiality). 
In CLMV07 these models were referred to as MOD1 and MOD1-bis;
the only difference between the two solutions was the maximum integration 
time permitted during the orbital integration, 
which was longer in MOD1-bis (5 Hubble times) than in MOD1 (2 Hubble times), 
i.e., MOD1-bis represents a more stationary solution. 

\section{The Self-Consistent Triaxial Models}
The Schwarzschild method used to build the self-consistent solutions
consisted in minimizing  the discrepancy between the model cell ``analytical'' masses (obtained by integration of a given $\rho(x,y,z))$ and the masses
given by a linear combination of orbits computed in the potential generated by $\rho$.
In practice, the quantities to be, separately,  minimized are :
\begin{eqnarray} \label{op_al_1}
\chi^2_{lum}=\frac{1}{N_{cells}} \sum_{j=1}^{N_{cells}}
\left(M_{j;lm}- \sum_{k=1}^{n_{orb}}C_{k;lm}B_{k,j;lm}\right)^2,~ 
\end{eqnarray}
and 
\begin{eqnarray} \label{op_al_2}
\chi^2_{dm}=\frac{1}{N_{cells}} \sum_{j=1}^{N_{cells}}\left(M_{j;dm}- 
\sum_{k=1}^{n_{orb}}C_{k;dm}B_{k,j;dm}\right)^2,~
 \end{eqnarray}
\noindent where $B_{k;j;lm(dm)}$ is the fraction of time that the $k$th orbit spends in the $j$th cell of
the luminous-matter grid (dark-matter grid); $M_{j;lm}$ is the mass which the model places in the $j$th
cell of the luminous-matter grid and $M_{j;dm}$ is the same quantity for the dark-matter grid.
$C_{k;lm}$ and $C_{k;dm}$    represent the total  mass, respectively, of 
luminous component and dark matter placed  over the $k$th orbit ($ 1\le k \le n_{orb}$).
The basic constraints are  $C_{k;lm}\ge 0$ and  $C_{k;dm} \ge 0$,
i.e., non-negative orbit weights.

The mass model considered  in CLMV07 for the luminous component was 
a triaxial generalizations of Dehnen's spherical model (Dehnen 1993) 
with a weak cusp. 
The density law for the luminous component was
\begin{equation}
\label{rhol} \displaystyle \rho_{l}(m) =\frac{M}{2 \pi
a_{l}b_{l}c_{l}} \frac{1}{m (1+m)^3}
\end{equation}
with
\begin{equation}
\label{m}\displaystyle
m^2=\frac{x^{2}}{a_{l}^{2}}+\frac{y^{2}}{b_{l}^{2}}+\frac{z^{2}}{c_{l}^{2}},
\qquad 0 < c_{l}< b_{l}< a_{l}
\end{equation}
and $M$ the total luminous mass. 

For the dark component the adopted mass density  was
\begin{equation} \label{den_dm}
\rho_{dm}(m')=\frac{\rho_{dm,0}}{(1+m')(1+{m'}^2)}
\end{equation}
with
\begin{equation} \label{ragg_ell_dm}
{m'}^2=\frac{x^2}{a_{dm}^2}+\frac{y^2}{b_{dm}^2}+\frac{z^2}{a_{dm}^2}
\end{equation}
and $\rho_{dm,0}$ is the central dark matter density (Burkert 1995).
Therefore the dark component  has a flat, low-density core. 
In the present work all quantities are given in units corresponding to $a_l=M_l=t_{cross}=1$ where $t_{cross}$ is the half mass crossing time of the system.
 Assuming $M_{l}=10^{11}M\odot$, $a_{l}=1$ kpc and $t_{cross}=26.5$ Myr.

\section{N-body initial conditions}
In order to study the dynamical properties  of MOD1 and MOD1-bis we sampled these models as $N$-body systems and let them evolve.
The initial conditions for the $N$-body integrations were set populating the $k$th orbit with a number of particle proportional to $C_k$
  and randomly choosing positions and velocities from the files containing the 
results of the orbital integration.

In this work we define $x$($z$)-tube orbits as orbits having a  
non-vanishing $x$($z$) component of the angular momentum.
All other orbits (either box or chaotic)  are classified  as 
\emph{semi-radial orbits}.
Regarding the initial conditions of the $N$-body systems, particular care should be given to the direction of rotation of the tube orbits. 
 In fact the persistence of the sense of motion  on tubes might be
cause of internal streaming motions, 
giving a  ``rotating'' model (Schwarzschild 1979; Merritt 1980).
As a consequence of the symmetries of the potential, the orbits  are invariant to a change in sign of the velocity 
and a decision should be made on the sense of motion of the particles placed on tube orbits. 
 To detect the effects of any rotational motion, and  to distinguish them from deformations due to dynamical instabilities, we set two
$N$-body simulations for each solutions(MOD1 and MOD1-bis) in the two extreme cases of high and ``zero'' angular momentum. 
In Table \ref{caratt}  some parameters of four $N$-body systems,  sampling MOD1 and MOD1-bis, are given. 

\begin{table}\small
\begin{center}
\begin{tabular}{|r|r|r|r|r|}
\hline
$MODEL$                & $Solution$     & $L$       &  $N_{lm}$  &  $N_{dm}$     \\
\hline
$HL$                & $MOD1$         & $23.71$    &  $19684$ &  $144886$   \\
\hline
$HL_{bis}$                & $MOD1-bis$     & $23.58$  &  $19709$ &  $146485$   \\
\hline
$LL$                & $MOD1$         & $0.40$ &  $19684$ &  $144886$   \\
\hline
$LL_{bis}$                & $MOD1-bis$     & $0.34$   &  $19709$   &  $146485$       \\
\hline

\end{tabular}
\caption{Initial features of  models used for the $N$-body simulations. $N_{lm(dm)}$ is the number of particles used to represent the
luminous(dark) component; $L$ is the absolute value of the total angular momentum.}\label{caratt}
\end{center}
\end{table}

\section{Results}
\begin{figure*}
\begin{center}
$\begin{array}{cc}
                \includegraphics[angle=270,width=3.2in]{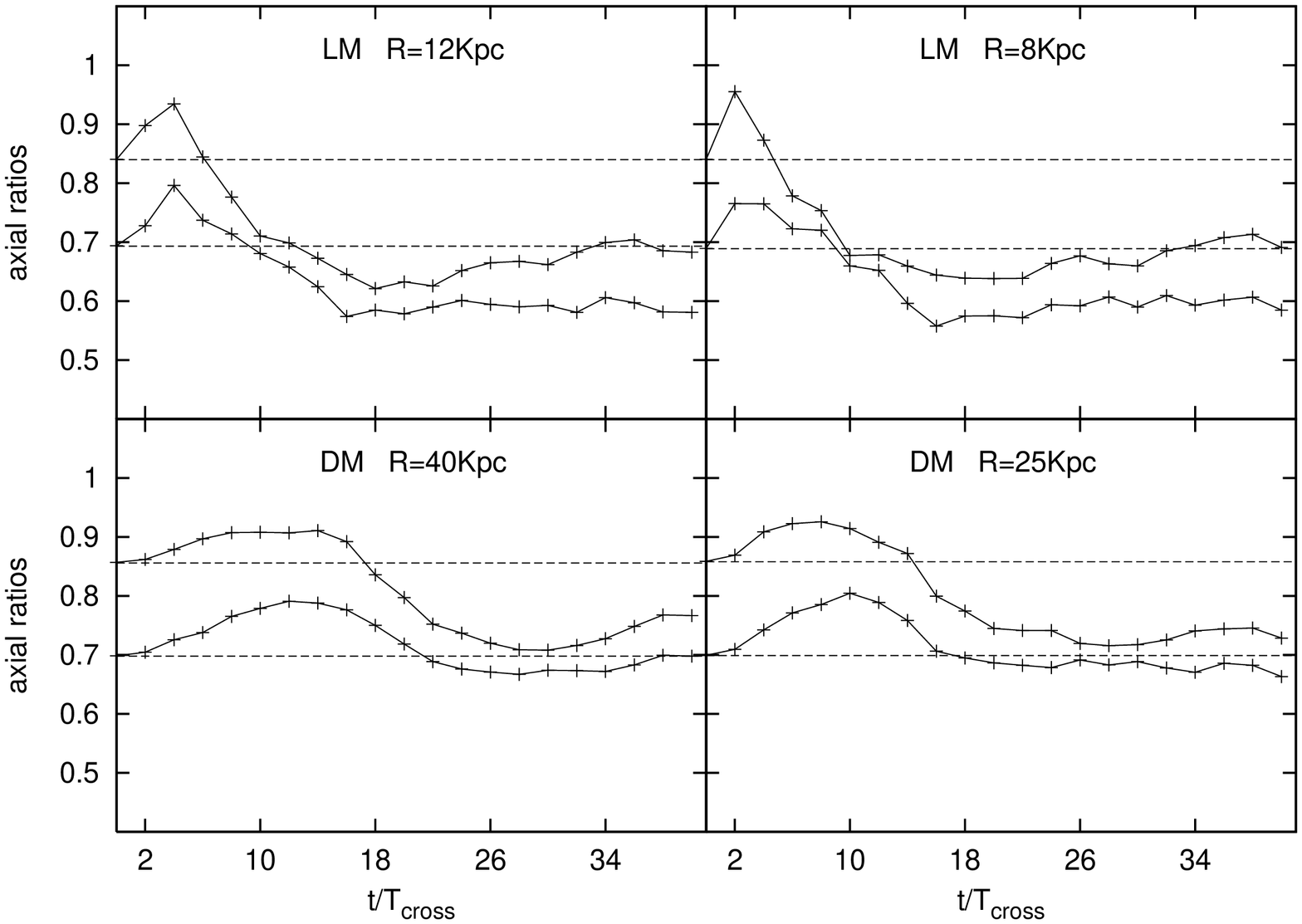}  & \includegraphics[angle=270,width=3.2in]{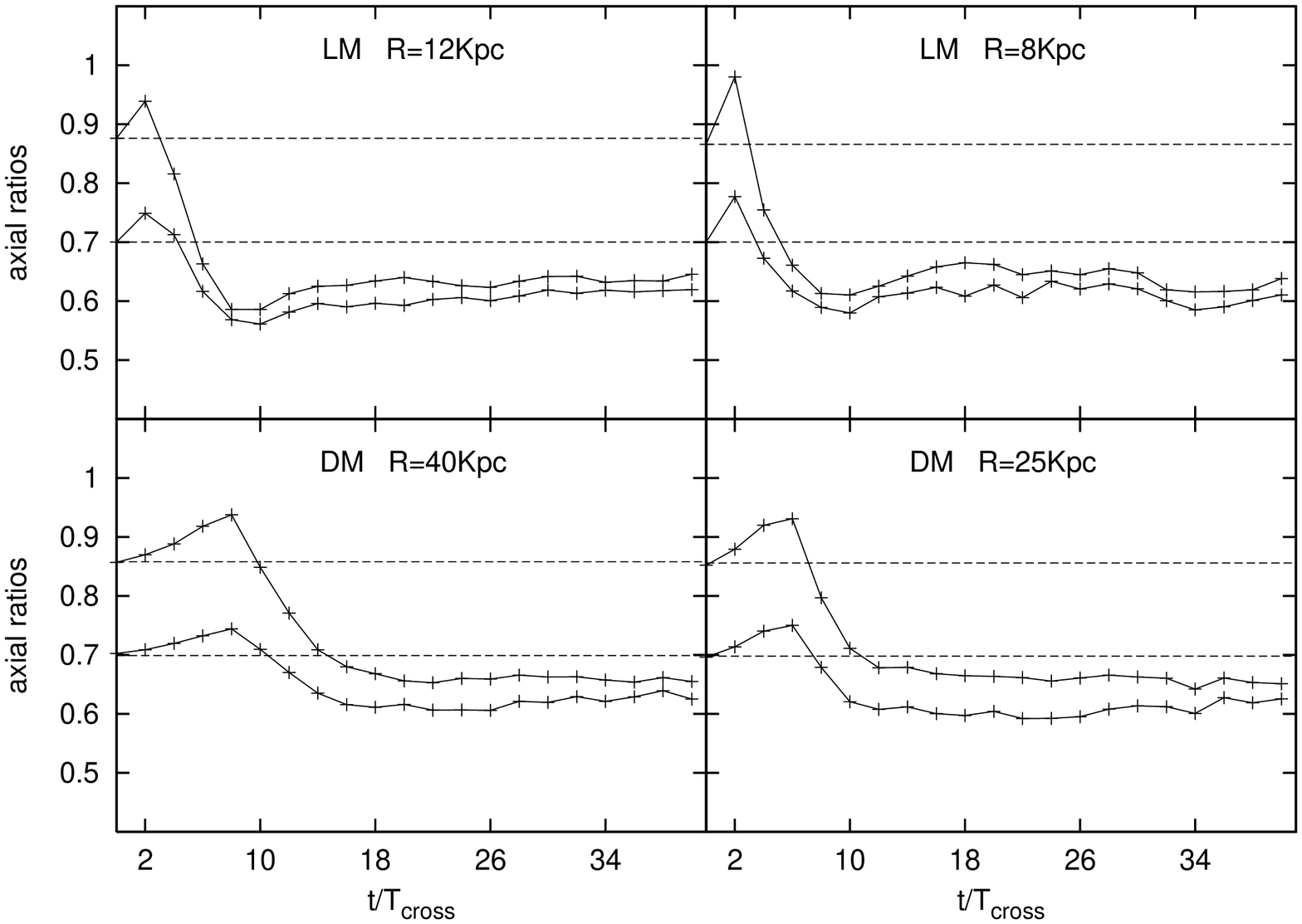} \\
                \includegraphics[angle=270,width=3.2in]{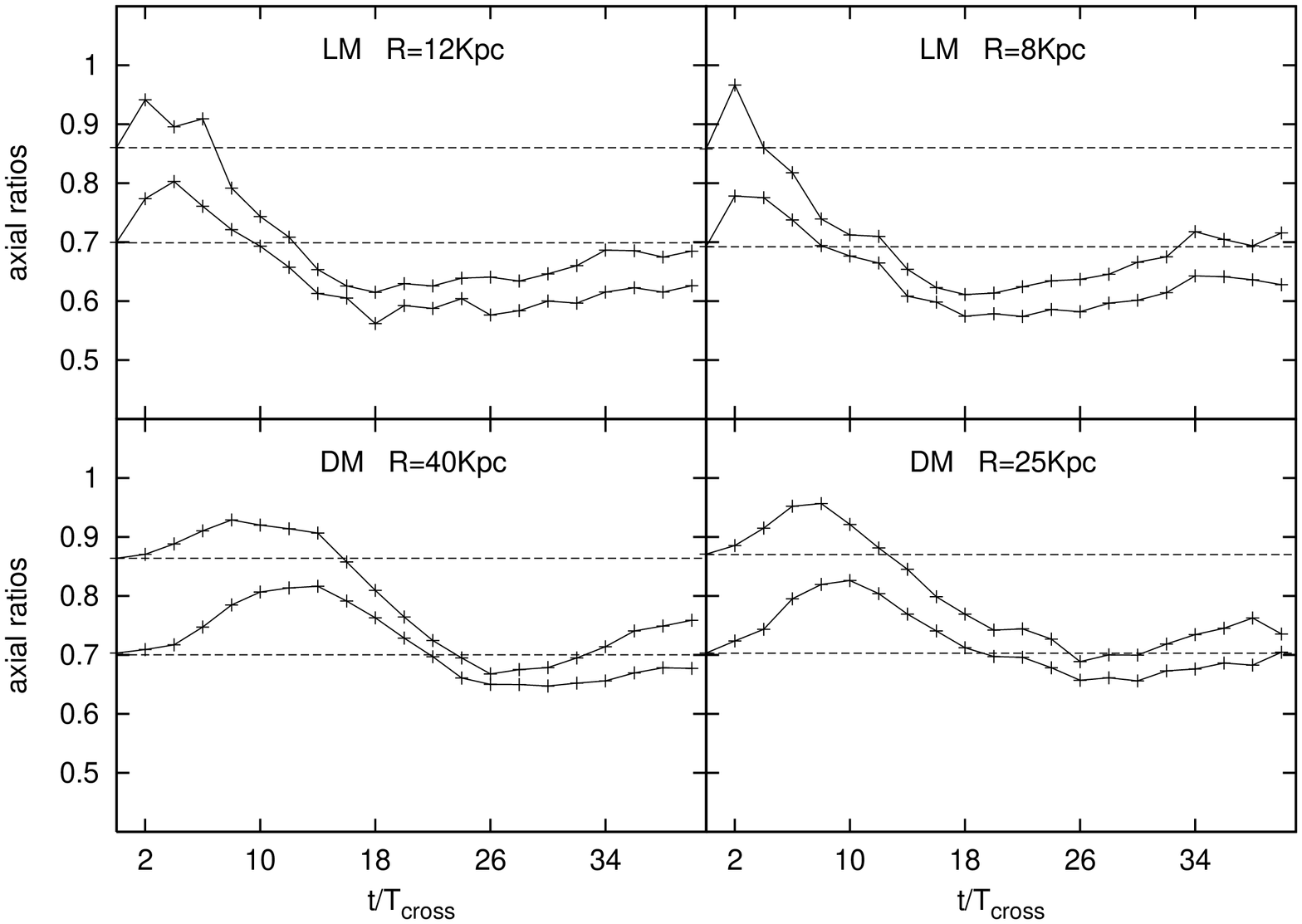} &  \includegraphics[angle=270,width=3.2in]{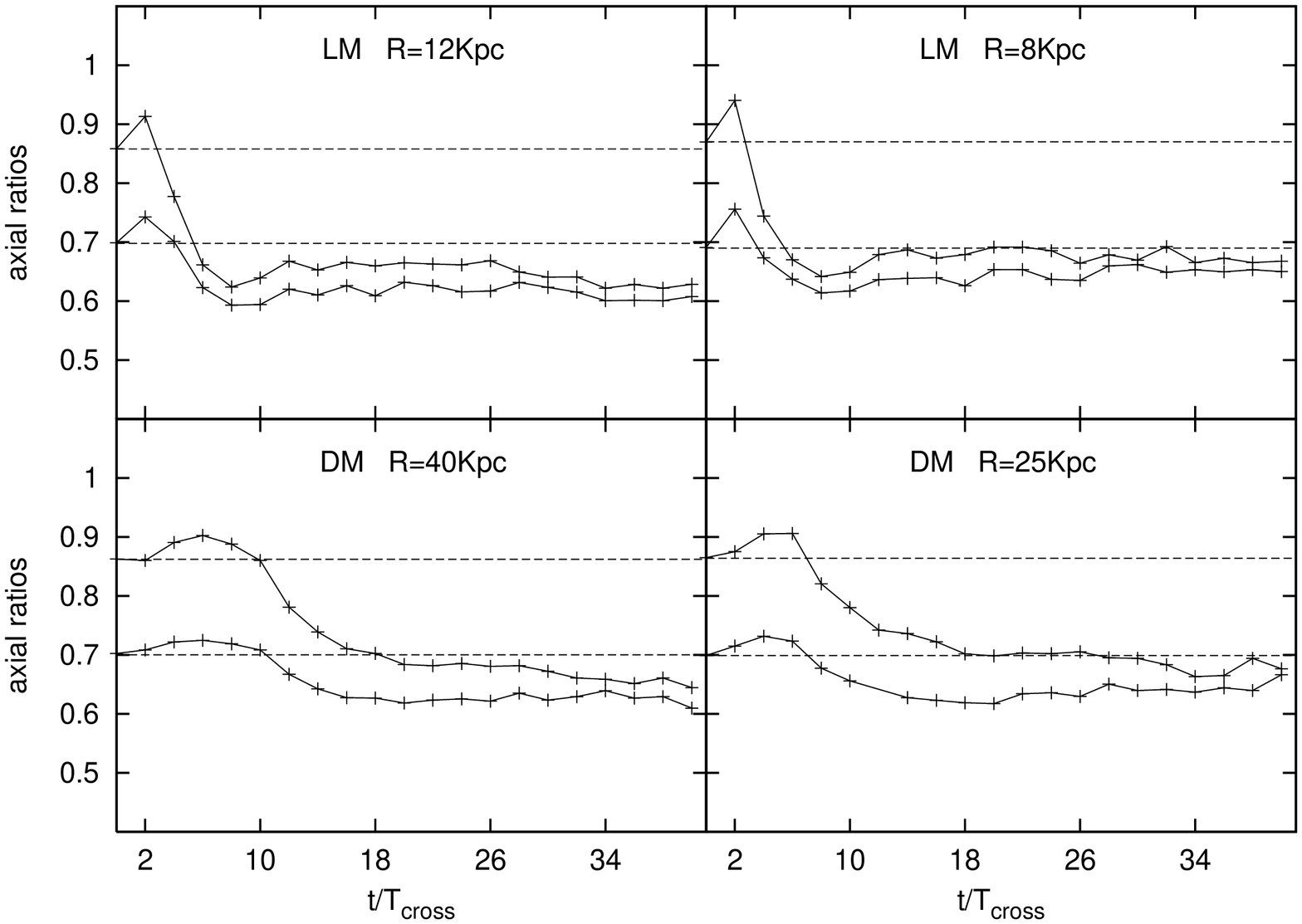}  
    \end{array}$\caption{\footnotesize{
Upper panels: evolution of the axial ratios of $HL$(left) and $LL$(right).
Lower panels: evolution of the axial ratios of $HL_{bis}$(left) and $LL_{bis}$(right).
R is the distance from the center where the axial ratios are evaluated. 
}}\label{RA1}
\end{center}
\end{figure*}

\begin{figure}
                \includegraphics[width=0.5\textwidth]{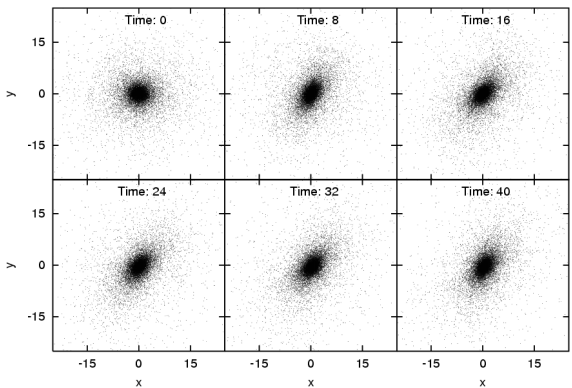} \\         
\caption{\footnotesize{Time evolution of the luminous matter 
for model $LL_{bis}$, projected onto the $x-y$ plane. }}
\label{SNAP}
\end{figure}

Simulations reveal that both  MOD1 and MOD1-bis evolve in shape, 
with no particular differences
detected between the two cases during the evolution. 
The variation of the axial ratios is shown in Fig. \ref{RA1}.
All the systems were found to have a  final quasi-axisymmetric prolate shape. 
As example, Fig. \ref{SNAP} shows snapshots of the
luminous component for model $LL_{bis}$.

We evaluated the anisotropy   parameter  $2T_r/T_t$ where $T_r=<v_{r}^2/2>$ and $T_t=<v_{t}^2/2>$.
For all $N$-body systems, roughlty the same values of the anisotropy 
parameters were found
 $[2T_r/T_t]_{dm}\approx 2$  and  $[2T_r/T_t]_{lm}\approx 1.4$. 
These values are quite high, in particular for the dark component, suggesting that a bias in the semi-radial orbits
 may have induced an analog to the radial-orbit instability (ROI)
of spherical models.

\section{Minimizing  the contribution from semi-radial orbits}
To show that the dynamical instability seen in the 
$N$-body simulations is due to the large radial ``pressure'' in the models, new solutions containing
a lower fraction of semi-radial orbits were computed by means of the orbital superposition technique and then evolved in time as $N$-body systems.
Our goal was to demonstrate the existence of a threshold value of the ratio between the number of tube orbits to that of semi-radial orbits above which stability is guaranteed. 

Following Poon and Merritt (2004), the relative orbital abundance  in the self-consistent models was changed by adding a new term 
in equations  (\ref{op_al_1}) and  (\ref{op_al_2}),   which become
\begin{eqnarray} \label{op_al_mod_1}
\chi^2_{lum}=\frac{1}{N_{cells}} \sum_{j=1}^{N_{cells}}
\left(M_{j;lm}- \sum_{k=1}^{n_{orb}}C_{k;lm}B_{k,j;lm}\right)^2 +{}
\nonumber\\
 +\sum_{k=1}^{n_{orb}}C_{k;lm}W_{k;lm}~~~~~~~~~~~~~~~~~
\end{eqnarray}
and 
\begin{eqnarray} \label{op_al__mod_2}
\chi^2_{dm}=\frac{1}{N_{cells}} \sum_{j=1}^{N_{cells}}
\left(M_{j;dm}- \sum_{k=1}^{n_{orb}}C_{k;dm}B_{k,j;dm}\right)^2+ {}
\nonumber\\
 +\sum_{k=1}^{n_{orb}}C_{k;dm}W_{k;dm}~~.~~~~~~~~~~~~~~~
\end{eqnarray}
\noindent
As above, the basic constraints were $C_{k;lm}\ge0$ and  $C_{k;dm}\ge0$ , i.e., non-negative orbit weights.

Here,  $W_{k;lm(dm)}$ is the ``penalty'' associated with the $k$th orbit of the luminous (dark) component, having the effect of filtering 
the orbital content in the solution: as $W_{k;lm(dm)}$ increases, the mass contribution $C_{k;lm(dm)}$ of the $k$th  orbit in the model  decreases.
We chose $W_{k;lm}= W_{k;dm}=0$ for the tube orbits and $W_{k;lm}\equiv W_{R;lm}>0$ and  $W_{k;dm} \equiv W_{R;dm}>0$ for the semi-radial orbits.
The new solutions were found by means of  the full set of orbits of the MOD1-bis catalog.
Four different $N$-body models were obtained by sampling new solutions, taking randomly the sense of rotation for the particles on tube orbits.
Each of these models was built using about $20000$  luminous and $150000$ dark matter particles.  
Table \ref{prop} gives  the anisotropy of these $N$-body systems: as expected, $[2T_r/T_t]_{lm}$ and  $[2T_r/T_t]_{dm}$  are  decreasing functions 
of $W_R$.

\begin{table}[b]\small
\begin{tabular}{|l|l|l|l|l|}
\hline
 MODEL           &$W_{R;lm}$  & $W_{R;dm}$   &  $[2T_r/T_t]_{lm}$  & $[2T_r/T_t]_{dm}$ \\
\hline
$A$              & $50$               & $50$               &      $0.512$    & $1.175$    \\
\hline
$B$              & $5$                & $5$                &     $ 0.784$   & $1.335$    \\
\hline
$C$              & $5\times 10^{-6}$  & $5\times 10^{-3}$  &     $1.220$  &   $1.754$   \\
\hline
$D$              & $5\times 10^{-6}$  & $50$               &     $ 1.230$    & $1.174$\\
\hline
\end{tabular}
\caption{Anisotropy parameters of the new models.}\label{prop}
\end{table}

\section{The new simulations}

\begin{figure*}
\begin{center}
$\begin{array}{cc}
                \includegraphics[angle=270,width=3.2in]{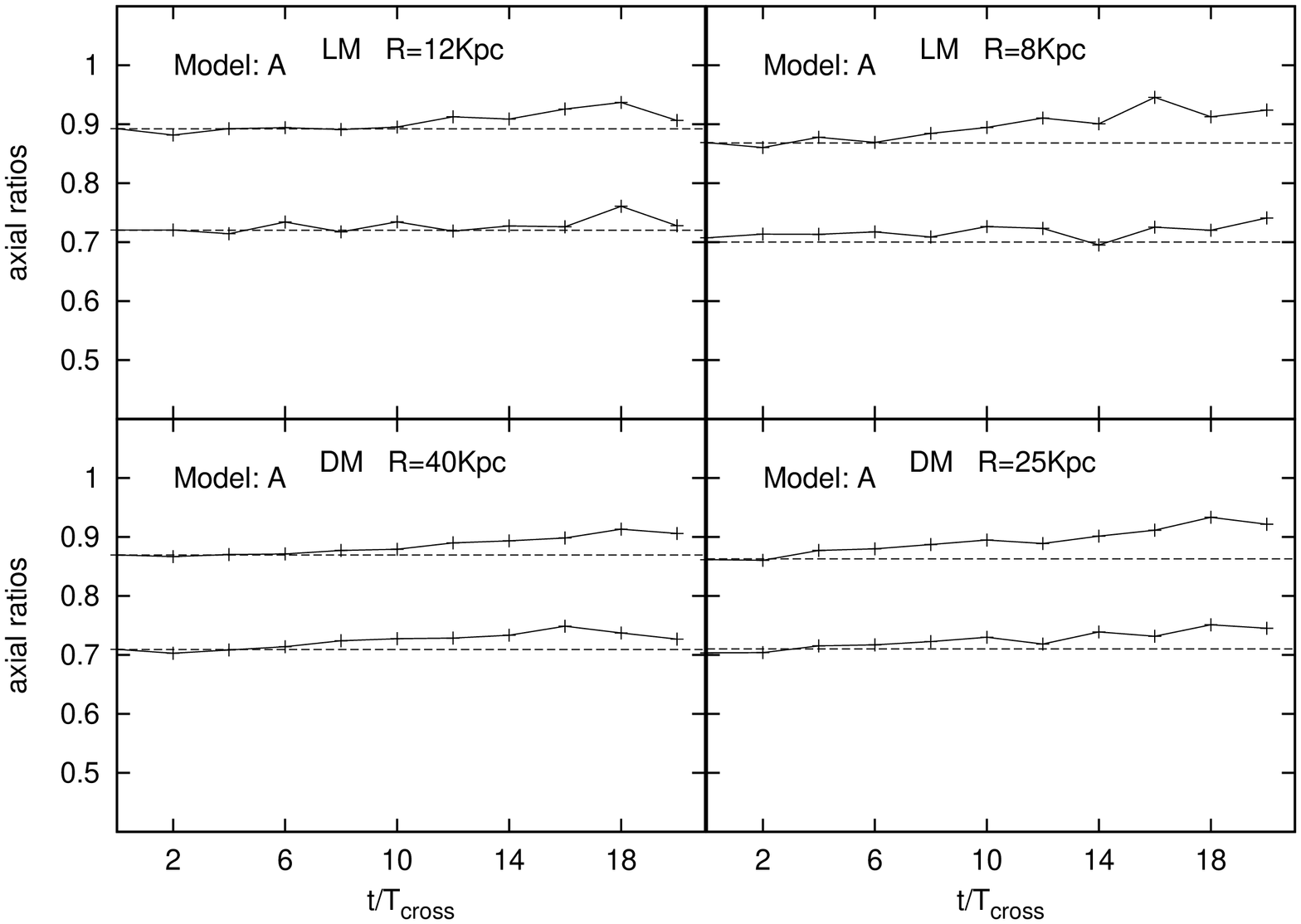}  &
                \includegraphics[angle=270,width=3.2in]{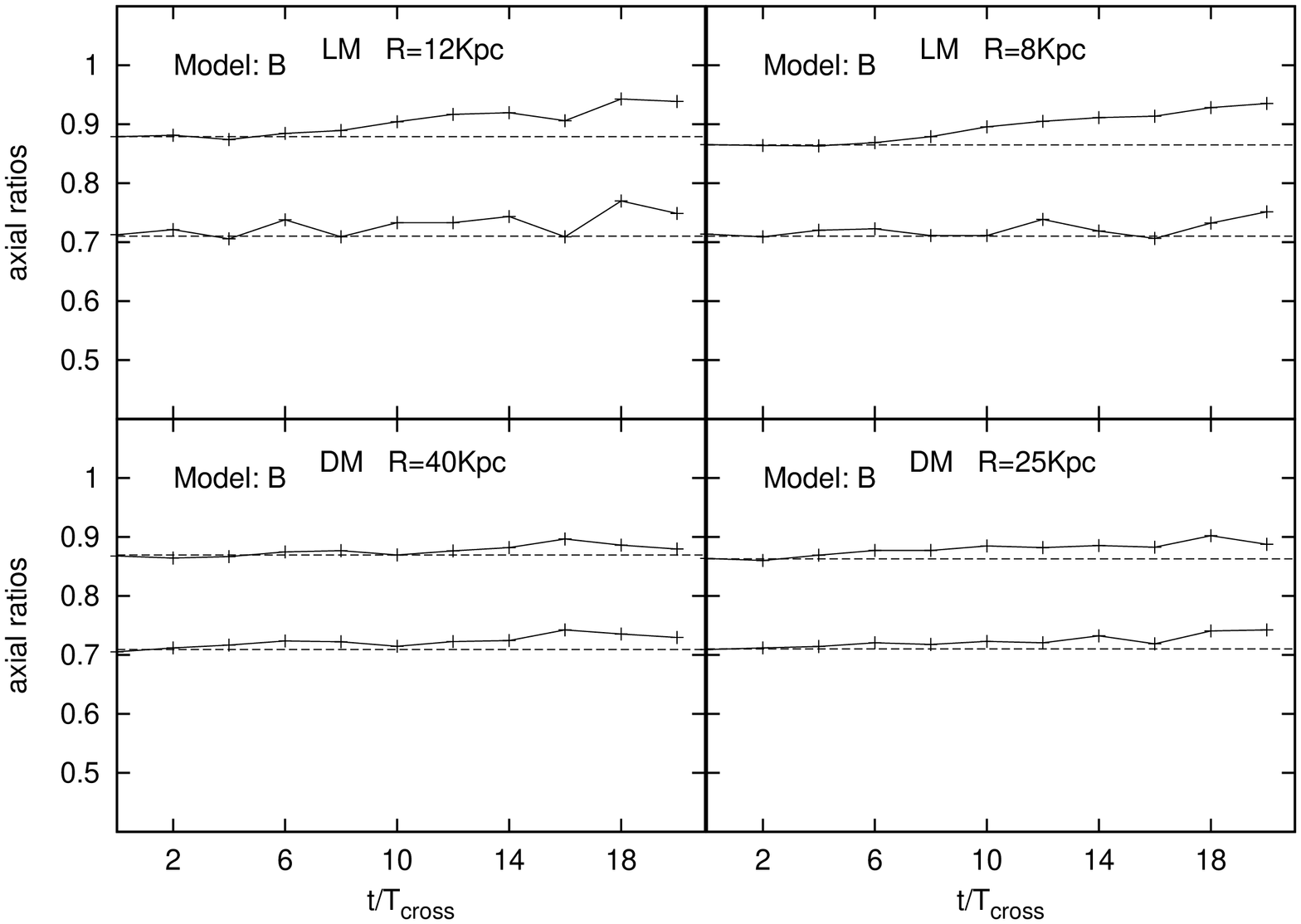}\\
                \includegraphics[angle=270,width=3.2in]{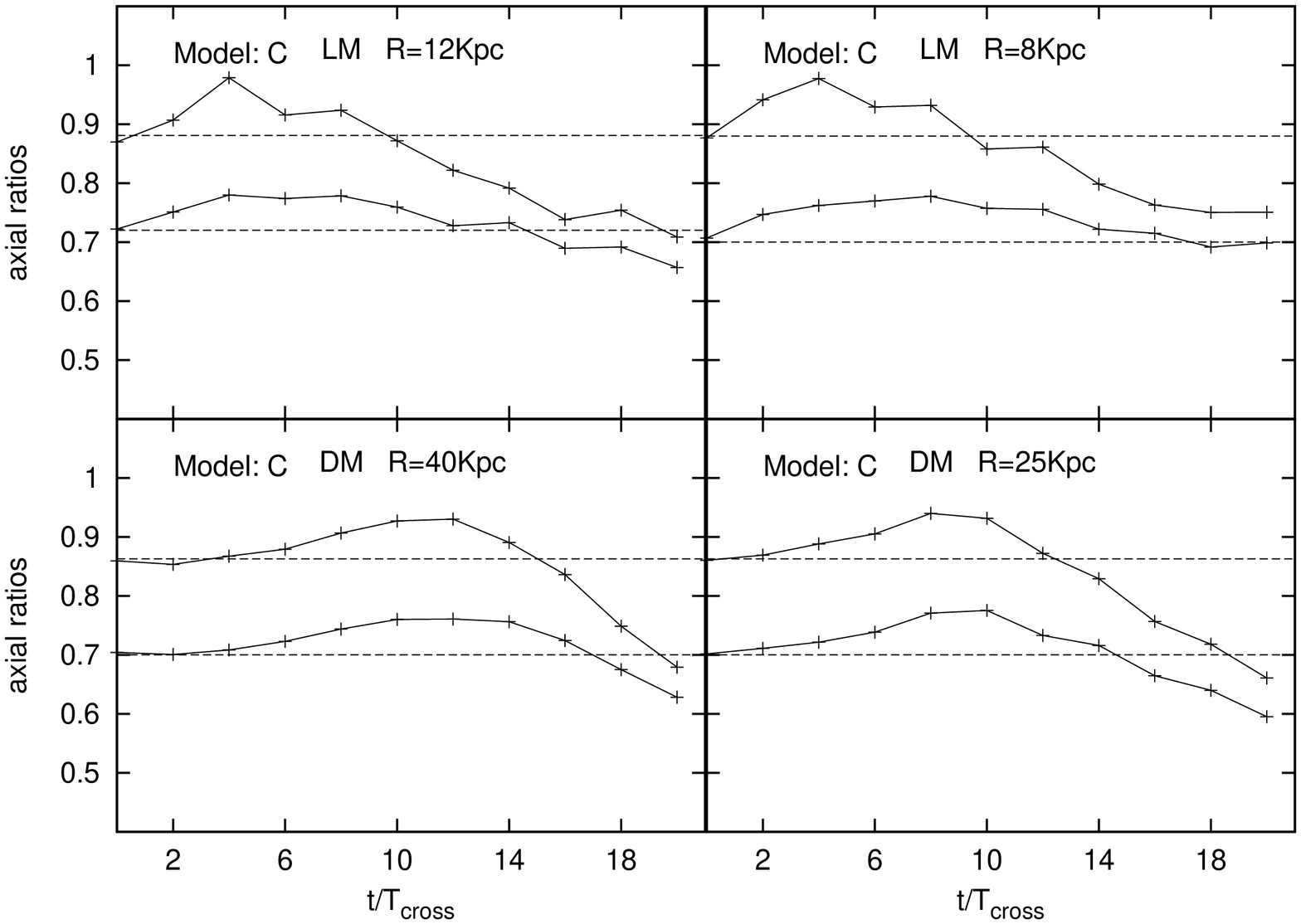}  &
                \includegraphics[angle=270,width=3.2in]{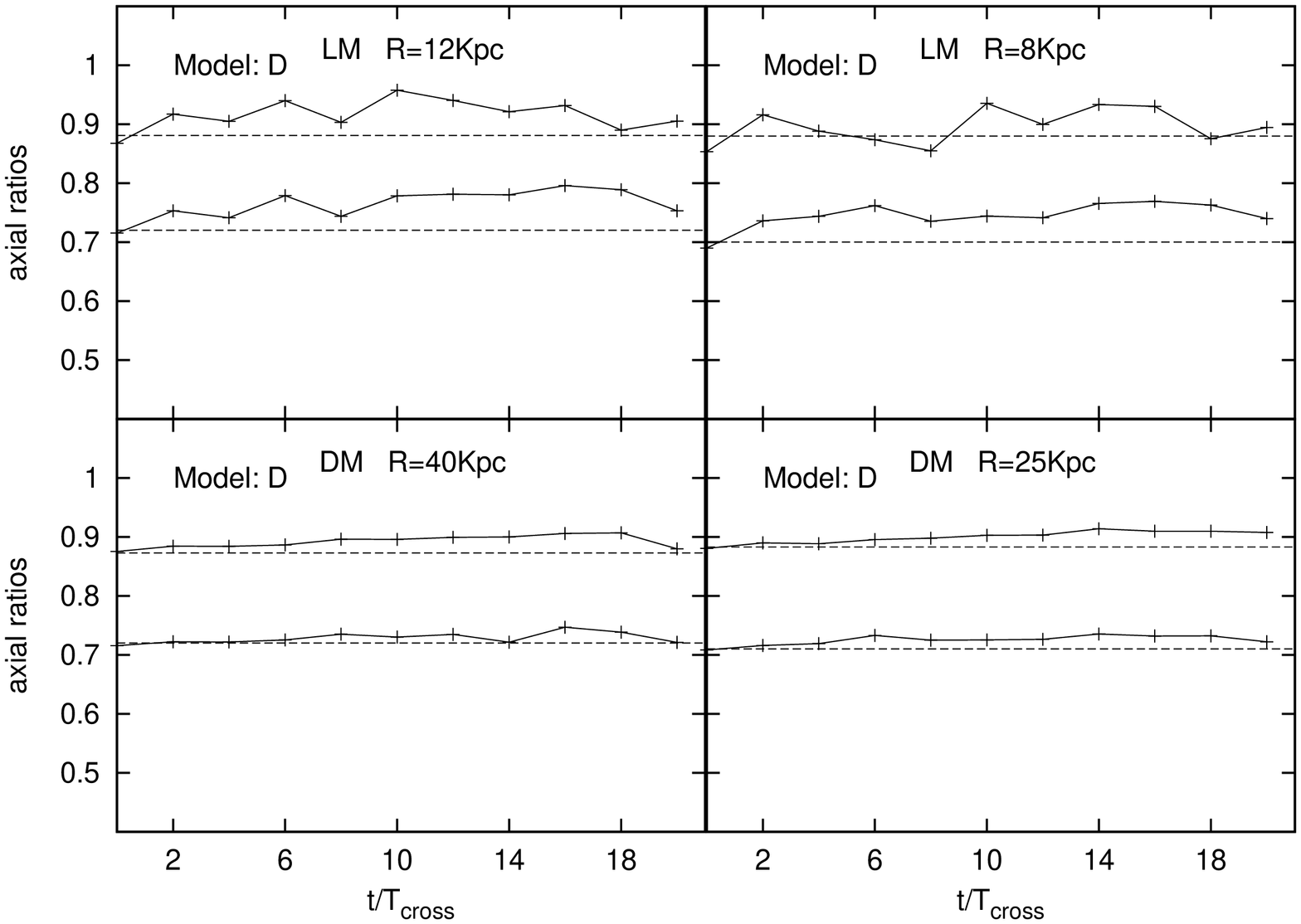} 
    \end{array}$
\caption{\footnotesize{
Evolution of the axis ratios for new models with lower number of semi-radial orbits. 
}}\label{RU1}
\end{center}
\end{figure*}

Fig. \ref{RU1} shows the axis-length evolution of the new models.
Model $C$ is clearly dynamically unstable, evolving into a prolate configuration. 
By contrast, for models $A$, $B$ and $D$, we did not observe the quick transition to instability that characterizes the evolution of models constructed in 
CLMV07.
Therefore, we conclude that  the instability can be suppressed when the  radial contribution to the solution is reduced.
The correlation between the bar formation and the value of the radial velocity dispersion
 is a clear sign that the dynamical instability, seen above, can be identified with the ROI.
Furthermore, the  results of model $D$  suggest that the instability disappears when  the dark halo gets more isotropic, i.e.,
the ROI  derives mainly by the initial high radial pressure in the dark matter halo. 

Finally,  we can draw the following conclusions from our analysis  of the various different runs:
\begin{itemize}
\item[i]The dynamical instability that characterizes the time evolution of the galaxy models built in CLMV07 disappears when solutions contain a
smaller number of semi-radial orbits;
therefore  it can be identified with the ROI;
\item[ii]The ROI is due to the high concentration of radially ``biased'' orbits in the \emph{dark matter halo}.
In particular numerical experiments suggest that  stable configurations are obtained when  $[2T_r/T_t]_{dm}<1.4$.
\end{itemize}

\section{Discussion}
Our study has important implications for future construction of self-consistent models of realistic galaxies.  
 In particular, it has been shown that the dynamical properties
of solutions cannot be directly deduced by the Schwarzschild method itself
since the temporal development of models can be strongly 
affected by the distributions in velocity space.
 In this context, $N$-body tests of the stability of the models seems 
to be indispensable tool for drawing conclusions about general properties of models constructed via orbital superposition. 

Regarding the ROI, because a complete understanding of its mechanism is still lacking, 
it would be interesting to  investigate more in depth its role  in the case of triaxial potentials.
Our simulations  show  dynamical features which cannot be detected in the case of spherical symmetry.
The method exploited here to change the semi-radial orbit contribution to the models may be useful for constructing other radially-unstable triaxial systems.

\end{document}